\documentclass[preprint,12pt]{aastex}
\usepackage{epsf}

\newcommand{\sig}{\:\lower0.6ex\hbox{$\stackrel{\textstyle >}{\sim}$}\:}
\newcommand{\sil}{\:\lower0.6ex\hbox{$\stackrel{\textstyle <}{\sim}$}\:}
\newcommand{\sigs}{\:\lower0.4ex\hbox{$\stackrel{\scriptstyle
      >}{\scriptstyle \sim}$}\,}
\newcommand{\sils}{\:\lower0.4ex\hbox{$\stackrel{\scriptstyle
      <}{\scriptstyle \sim}$}\,}

\slugcomment{\sf (ApJ, in press)}

\shorttitle{Draco: Not A Tidal Dwarf}
\shortauthors{Klessen, Grebel, \& Harbeck}

\begin{document}

\title{ Draco -- A Failure of the Tidal Model } 
\author{Ralf S.\ Klessen\altaffilmark{1,2}, Eva K. Grebel\altaffilmark{3},
and Daniel Harbeck\altaffilmark{3}}

\altaffiltext{1}{Astrophysikalisches Institut Potsdam, An der
 Sternwarte 16, D-14482 Potsdam, Germany ({\tt rklessen@aip.de})}
\altaffiltext{2}{UCO/Lick Observatory, University of California at Santa Cruz, 
 Santa Cruz, CA 95064, U.S.A. ({\tt ralf@ucolick.org})} 
\altaffiltext{3}{Max-Planck-Institut f{\"u}r
 Astronomie, K{\"o}nigstuhl 17, D-69117 Heidelberg, Germany
({\tt grebel@mpia.de, dharbeck@mpia.de})}

\begin{abstract}
We test whether the structural properties of the 
nearby dwarf spheroidal (dSph) galaxy Draco, a well-studied
Milky Way companion, can be reconciled with the suggestion
that dSphs are unbound tidal remnants with a large depth
extent along the line of sight.  In order to apply the observational
test of this hypothesis
suggested by Klessen \& Zhao, we use public photometric
data from the Sloan Digital Sky Survey (SDSS) to explore the
width of Draco's blue horizontal branch over a range of areas covering
0.06 square degrees to 6.25 square degrees centered on Draco.  
The SDSS database is the only currently existing database with
sufficient depth and area coverage to permit a stringent test
of the tidal models.
Blue horizontal branch stars were chosen as tracers of Draco's 
spatial extent and depth due to their low contamination
by Galactic foreground stars and since they have a spatially more
extended distribution than the more centrally concentrated red
horizontal branch stars.  Indeed, we show that blue horizontal branch stars
extend beyond the previously inferred limiting radii of Draco,
consistent with the observed absence of a truncated stellar surface
density profile of this dSph (Odenkirchen et al.).
Following the method of Klessen \& 
Kroupa, we calculate new models for a galaxy without dark
matter, using Draco's morphological properties as constraints.
The resulting models are unable to reproduce the narrow observed
horizontal branch width of Draco, which stays roughly constant
regardless of the field of view.
We conclude that Draco cannot be the remnant of a tidally
disrupted satellite, but is probably strongly dark-matter
dominated,
as suggested independently by the structural analysis conducted
by Odenkirchen et al.\ and by the kinematic analysis of
Kleyna et al.
\end{abstract}

\keywords{galaxies: dwarfs --- galaxies: evolution --- galaxies: kinematics and
dynamics  --- galaxies: structure --- Local Group}

\section{Introduction}
\label{sec:intro}
Dwarf spheroidal (dSph) galaxies are very unusual astrophysical
objects. Their stellar masses are comparable to those of massive
globular clusters (Harris 1996), but unlike globular clusters, which
are dense and compact, dSphs are very extended and diffuse
assemblies of stars. Their limiting radii exceed those of globular
clusters by several hundreds (Irwin \& Hatzidimitriou 1995) and 
reach several kpc, while the
velocity dispersion of their stars again is comparable to what is 
measured in globular clusters. This makes them unusual in the sense
that if dSphs were bound objects in virial equilibrium, then their
gravitational mass as implied by the velocity dispersion must exceed
the luminous mass observed in stars by one to two orders of magnitude. 
Dwarf spheroidal 
galaxies therefore are often claimed to be completely dominated by
invisible dark matter (see, e.g., the review on that subject by Mateo
1997).

The assumption of virial equilibrium, however, may be challenged since
many dSphs orbit massive $L_*$-type galaxies like our Milky Way and
are therefore subject to strong tidal torques. This influences their
internal dynamics, and may bring them on the brink of disruption as is
claimed for the nearby dSph Ursa Minor (Mart\'{\i}nez-Delgado et al.\ 
2001) or even complete dissolution as observed in the case of the
Sagittarius dwarf galaxy (Ibata, Gilmore, \& Irwin 1994). It is
therefore a valid approach to hypothesize that many, if not all, dSph
companions of the Milky Way may be the remnants of tidally disrupted
satellites without appreciable amounts of dark matter.

The `tidal scenario' has received considerable attention.  In early
studies, Kuhn \& Miller (1989) concentrated on the effects of resonant
coupling between orbital and internal motions, while Kuhn (1993)
investigated diffusion in phase and configuration space of assemblies
of stars orbiting various rigid potentials. Both studies concluded
that indeed tidal effects could mimic the observed large mass-to-light
ratios. The opposite conclusion was derived by Oh Lin, \& Aarseth
(1995) who included the effects of the satellites' self-gravity into
their studies, and by Piatek \& Pryor (1995) who focused on tidal
effects during the satellites' perigalactic passage. Further discussion
on tidal mass loss can be found in Helmi \& White (1999), Johnston,
Sigurdsson, \& Hernquist (1999), Bekki, Couch, \& Drinkwater (2001).

The most detailed models of the long-term evolution of low-mass
satellite galaxies in the tidal field of the Milky Way have been
discussed by Kroupa (1997) and Klessen \& Kroupa (1998) who
demonstrated that a remnant containing a few percent of the initial
mass prevails as a long-lived and distinguishable entity for a period
of several billion years after the initial satellite dissolves. They
proposed that what appears to be a bound dSph galaxy to a terrestrial
observer may in fact be the unbound remnant of a tidally disrupted
satellite galaxy on an eccentric orbit with two tidal arms extending
along a small angle from the line of sight of a terrestrial observer.
This model successfully demonstrates that high velocity dispersions in
dSphs may be obtained without internal dark matter.  It also accounts
for the distorted morphology of, e.g., the double-peaked dSph Ursa
Minor (e.g., Kleyna et al.\ 1998). The tidal model furthermore
predicts an appreciable spread in stellar distance moduli as the tidal
remnant will have considerable depth along the observer's line of
sight, which is an observationally testable prediction. Another model
prediction is a radial velocity gradient during most of the evolution
after tidal disintegration. These effects have been studied in detail
by Klessen \& Zhao (2002), who suggested to use the width of the
horizontal branch (HB) in the color-magnitude diagram (CMD) as
possible test for the tidal origin of the Galactic dSphs (see also
Klessen \& Kroupa 1998).

In our current investigation we apply the HB test to the nearby dSph
galaxy in the constellation of Draco.  Our goal is to construct a
tidal model for Draco that complies with {\em all} observational
properties of the galaxy known to date.  We will see that this attempt
fails, and we thus claim that Draco cannot be of tidal origin. This
implies that either Draco is strongly dark-matter dominated, or,
alternatively, that modifications to the law of gravity are necessary
(Milgrom 1983, 1995; see Sanders \& McGaugh 2002  for a review).  We focus on
Draco, because it is particularly well suited for testing the tidal
model.  Draco is the only dSph to-date for which high-quality, deep,
multi-color CCD imaging exists covering both Draco as well as a seven
times larger region than its tidal radius.  Draco is a close satellite
galaxy of our Milky Way.  Its compact and seemingly undistorted
morphological appearance (Odenkirchen et al.\ 2001a) as well as its
thin HB in the CMD can be used to define stringent constraints on
tidal models.

Our line of reasoning is as follows. First, in Section
\ref{sec:draco}, we discuss morphological and kinematic properties of
Draco.  We use public data from the Sloan Digital Sky Survey (SDSS,
see York et al.\ 2000; Gunn et al.\ 1998; and Stoughton et al.\ 2002
for more information) to determine the width of the HB of Draco. In
Section \ref{sec:model}, we attempt to construct an appropriate model
for the tidal origin of Draco. Our finding is that this is not
possible. Finally, in Section \ref{sec:summary} we summarize and
discuss our results.

\section{Morphology and Horizontal-Branch Thickness of the Draco Dwarf
Spheroidal Galaxy}
\label{sec:draco}

If dSphs are unbound tidal remnants, then nearby dSphs provide the best
test objects since here we can study the then expected depth effects
with the greatest precision.
Draco is a nearby Milky Way satellite at a distance of only $\sim$ 80
kpc.  With an absolute magnitude of $M_V = -9.4$ mag, a central
surface brightness of $\mu_V = 24.4 \pm 0.5$ mag arcsec$^{-2}$, and
lack of detectable gas (M$_{HI} < 8000$ M$_{\odot}$), it is a typical 
low-luminosity Local Group dSph (see Grebel 2000 and
Grebel, Gallagher, \& Harbeck 
2003 for a recent compilation of Local Group dSph properties).  Draco
is dominated by metal-poor stellar populations older than 10 Gyr
and possesses a well-populated HB (e.g., 
Stetson, McClure, \& VandenBerg 1985; Grillmair et al.\ 1998; 
Aparicio, Carrera, \& Mart\'{\i}nez-Delgado 2001).  The metallicity
spread observed in Draco (e.g., 
Carney \& Seitzer 1986; Shetrone, C\^ot\'e, \& Sargent 2001)
shows that its early star formation history was extended and complex,
unlike that of a globular cluster.   

\subsection{Morphological Properties of Draco}
\label{subsec:morphology-draco}

Draco and its surroundings have been mapped in their entirety 
with deep, homogeneous five-color photometry by the SDSS.
The SDSS obtains CCD imaging in the passbands {\em u}, {\em g}, {\em r}, {\em i},
and {\em z} of a contiguous region of $< 10^4$
deg$^2$ centered on the north Galactic cap.  The public data presented
here are part of the SDSS Early Data Release (EDR, Stoughton
et al.\ 2002).  The SDSS multi-color photometric database permits one 
to employ optimized filtering techniques to enhance the signal of the 
stellar population associated with the target of interest versus
contaminants such as foreground stars, and to
remove background galaxies based on shape parameters (see
Odenkirchen et al.\ 2001a,b for details).  The SDSS also provides
corrections for Galactic foreground 
extinction using Schlegel, Finkbeiner, \& Davis' (1998) maps.
Odenkirchen et al.\ (2001a) carried out a morphological study of
Draco based on these SDSS data covering more than 27 deg$^2$
of the galaxy itself and its surroundings, extending earlier 
photographic work by Irwin
\& Hatzidimitriou (1995).  Odenkirchen et al.\ found Draco to 
exhibit a flattened stellar surface density distribution with
constant ellipticity ($e = 0.29 \pm 0.02$, position angle $=
88\degr \pm 3\degr$).  The limiting radius of Draco is 
$40.1'$ along the major axis
from an empirical King profile (King 1962), and
$49.5'$ from a theoretical King profile (King 1966), which is 41\% to
75\% more extended than found in the earlier photographic study
of Irwin \& Hatzidimitriou (1995).  
Overdensities consistent with the presence of extratidal
stars were not detected down to a level of $10^{-3}$ of 
Draco's central surface density or to a surface
brightness of $\sim 33.5$ mag arcsec$^{-2}$.  
Odenkirchen et al.\
point out that that Draco's density profile may be equally well
fitted by a S\'ersic profile with an exponent of $n=1.2$, since
there is no detectable break in the profile, which may indicate
an even more extended distribution below the current detection limit.
The isopleths of Draco's stellar surface density
are highly regular, nested ellipses that retain the same 
position angle with increasing distance from the dSph's center.
These properties led Odenkirchen et al.\ to infer that Draco
is bound and strongly dark-matter dominated.  

Aparicio et al.\ (2001) used B and R CCD photometry covering 1 square
degree including parts of Draco and some of its surroundings. 
They also found a large tidal radius ($41'$) and no evidence for 
extratidal stars.  Another structural study of mostly non-contiguous
CCD imaging in Draco and surroundings covering a total area of
$\sim 1.36$ square degrees in V and R
was carried out by Piatek et al.\ (2002),
and again confirmed the lack of extratidal stars.  A recent CCD study of
the central part of Draco (0.17 deg$^{-2}$; Bellazzini et al.\ 2002) 
using the V and I filters 
also found the smooth elliptical isopleths presented by Odenkirchen
et al.\ (2001a).
Aparicio et al.\ (2001) compared Draco's CMD to 
one of the models proposed by Kroupa (1997), RS1-4,
according to which Draco is extended along the line of sight.
Using the width of the HB of Draco, particularly
of the red HB, 
they found an upper limit of $\sim 14$ kpc for Draco's depth extent, which
is at most one third of the extent predicted by Kroupa's model.
Aparicio et al.\ conclude that their upper limit precludes the explanation
of Draco's velocity dispersion as caused entirely by projection effects,
and that dark matter is a more likely culprit.

Finally, Kleyna et al.\ (2001, 2002) studied the radial velocity dispersion
of Draco's red giants as a function of projected distance out to a distance
of $\sim 25'$ from Draco's center.  {From} this inner region
they infer that Draco has a gradually
rising rotation curve and is embedded in an extended dark matter halo rather
than a halo in which mass follows light.  Kleyna et al.'s (2001) 
calculations suggest that Draco is close to its perigalacticon and that 
its tidal radius exceeds $1^{\circ}$.

\subsection{The Horizontal Branch of Draco}
\label{subsec:HB-draco}

Taking advantage of the large area coverage afforded by the SDSS data,
we investigate the distribution of HB stars in Draco
and surroundings.  As compared to conventional CCD imaging data, the
SDSS data allow us to study the entire projected area within which
models of the kind calculated by Kroupa (1997), Klessen \& Kroupa
(1998), Klessen \& Zhao (2002), and in the current paper predict 
stars originating from a disrupted dSph galaxy without dark matter.  
Furthermore, the SDSS database allows us to exclude objects 
identified as galaxies
or as quasars based on their shape parameters or location in multi-color space
(see, e.g., Richards et al.\ 2001; Stoughton et al.\ 2002).  

\begin{figure}[th]
\unitlength1cm
\begin{center}
\begin{picture}(11.0, 9.0)
\put(-1.0,-0.7){\epsfxsize=13cm\epsfbox{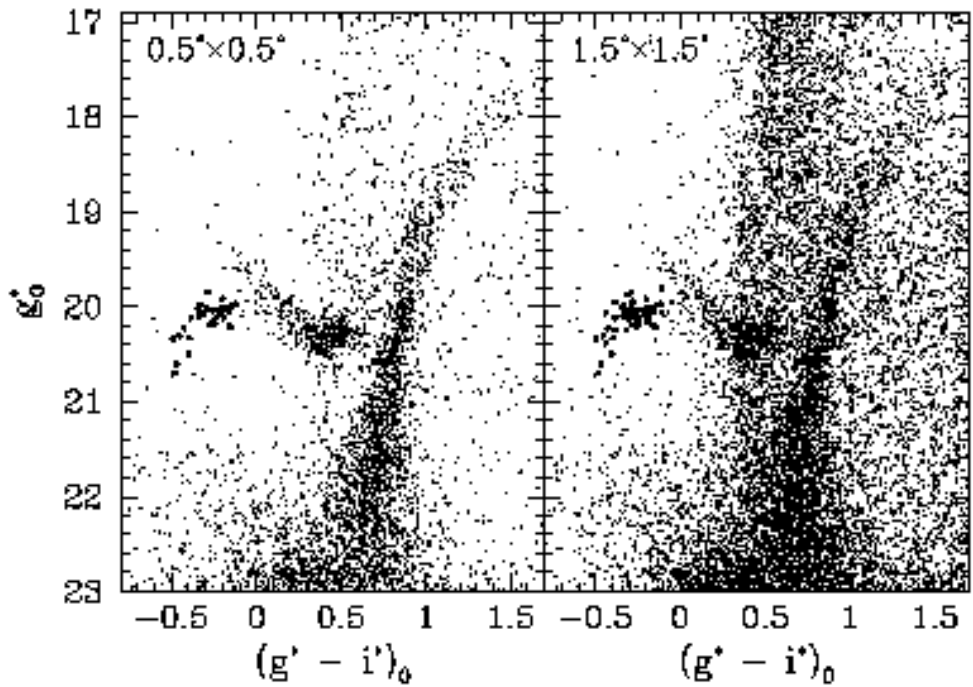}}
\end{picture}
\end{center}
\caption{\small Color-magnitude diagrams of Draco and surroundings.
The magnitudes were dereddened using the extinction maps of
Schlegel et al.\ (1998).  The left panel shows all stars within
a region of $0.5^{\circ} \times 0.5^{\circ}$ centered on Draco,
while the right panel contains all stars within a 
$1.5^{\circ} \times 1.5^{\circ}$ area.  In the left panel, Draco's
red giant branch, prominent red horizontal branch and diagonal
RR Lyrae strip stand out.  Blue horizontal branch candidates 
fulfilling our selection criteria (Section \ref{subsec:HB-draco})
are marked by fat dots.  In the right panel, the increasing 
contamination by Galactic foreground stars, which overlap in
particular with Draco's red horizontal branch stars, is obvious.
In comparison, the blue horizontal branch  locus suffers little
contamination. \label{Draco_sample_CMD}}
\end{figure}

The red HB of Draco is heavily contaminated by Galactic
foreground stars (Fig.\ \ref{Draco_sample_CMD}), particularly by the 
metal-weak Galactic halo main sequence
and, to a lesser extent, by main-sequence stars from the thick disk
(Chen et al.\ 2001).  While we can easily perform statistical field
star subtractions in the color-luminosity range of Draco's HB, 
this contamination hampers our ability to determine the {\em
width} of the red HB.  This problem becomes particularly
severe at larger distances from the center of Draco, where the stellar
density of Draco drops rapidly.  We therefore decided not to 
include red HB stars in our analysis.

Apart from a well-populated red HB, Draco contains a 
sizeable number of RR Lyrae stars, which overlap with the red
HB locus and become, on average, increasingly more 
luminous toward bluer colors (e.g., Bellazzini et al.\ 2002; their
Figure 6).  To avoid uncertainties in luminosity (and hence width)
introduced by RR Lyrae variables, we excluded their loci from our 
analysis as well.

This leaves us with the blue HB.  While Draco's blue
HB is sparsely populated in comparison to the red
HB, blue stars generally have the advantage of being easily
recognizable even in regions where Galactic field stars dominate
otherwise,
since Galactic field star contamination is very low at these blue colors.
The field star contribution in the area of the red
HB amounts to $62.2\pm5.3$ objects per square degree,
while for the blue HB it is only $1.8\pm0.9$ deg$^{-2}$.
Some of the latter contaminants
may be unidentified quasars at redshifts between
$\sim 2$ and $\sim 3$ (e.g., Fan 1999; Richards et al.\ 2001).

\begin{figure}
\unitlength1cm
\begin{center}
\begin{picture}(11.0, 10.7)
\put(-1.5,-1.5){\epsfxsize=13cm\epsfbox{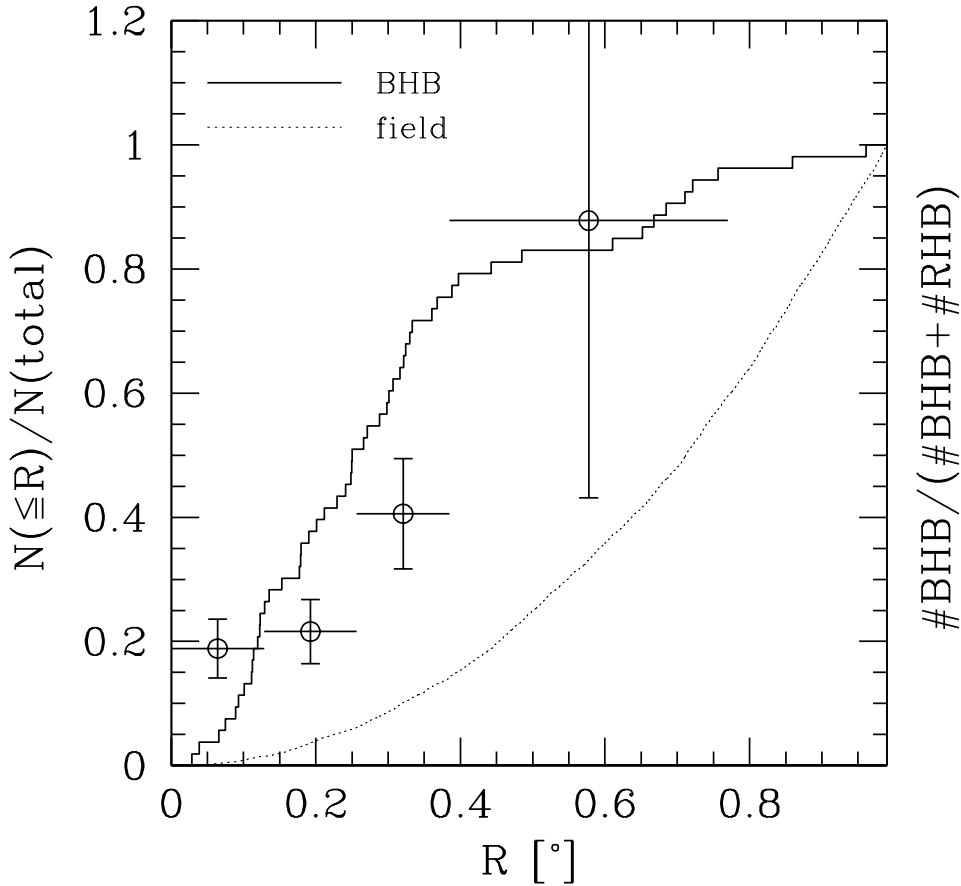}}
\end{picture}
\end{center}
%\plotone{f2.eps}
\caption{\small Cumulative radial distribution of blue horizontal branch
(BHB, solid line) stars.  The label and numbers of the left-hand side
refer to the number of stars within the radial distance R from the galaxy
center, N($\le$R), normalized by the total number of stars, N(total).
For comparison, 
the dotted line indicates the distribution of Galactic thin  disk stars
as a homogeneously distributed, control field population unrelated to Draco
(see Chen et al.\ 2001 for details).  
Open circles indicate the fraction of BHB stars 
with respect to the total number of horizontal branch stars 
(\#BHB / (\#BHB + \#RHB)).  The fraction
of BHB stars increases with increasing projected
angular distance from the center of Draco.
The inner three open circles sample the BHB fraction within one, two,
and three core radii, whereas the outermost bin combines three bins
of this size for improved number statistics.  The bin width is indicated
by the horizontal bars.
The vertical bars show the uncertainty of (\#BHB / (\#BHB + \#RHB)).
The BHB stars can be traced out to at least the limiting radius of Draco
inferred from using King (1966) models (Odenkirchen et al.\ 2001a).
Draco's BHB stars show
a more extended distribution than the more centrally concentrated
red horizontal branch stars.  This kind of population gradient among
old stars is a fairly common feature in dwarf spheroidal galaxies
(see Harbeck et al.\ 2001) and underlines the suitability of blue
horizontal branch stars as tracers of the spatial extent of dwarf
galaxies.  
\label{Draco_HB_cumul}
}
\end{figure}

Many dwarf galaxies show population gradients in the 
sense that younger populations are more centrally concentrated than
older ones (see Grebel 1997, 1999, 2001 for reviews).  Gradients may even
exist in the old populations of dwarf galaxies:  Harbeck et al.\
(2001) showed that in dwarf galaxies blue HB 
stars often have a spatially more extended distribution than red HB 
stars. This ``second parameter'' variation may be caused by
age or metallicity, or both, or additional factors (e.g.,
Hurley-Keller, Mateo, \& Grebel 1999; Harbeck et al.\ 2001).  
In Draco, our analysis of the SDSS data reveals that Draco's HB stars follow
the same trend.  Beyond galactocentric distances $> 0.5^{\circ}$
(Figure \ref{Draco_HB_cumul}), the contribution of red HB stars 
decreases:  the fraction of blue HB 
stars (measured as the number of blue HB stars, \#BHB,
over the sum of \#BHB and the number of red HB stars,
\#RHB) increases from $< 20$\% in the inner region of Draco
to $\sim 90$\% in its outskirts 
(Figure \ref{Draco_HB_cumul}).  This makes blue HB 
stars useful tracers of the spatial extent of Draco.

%As emphasized by Odenkirchen et al.,
%the absence of a truncated stellar density profile of Draco may 
%indicate an even more extended distribution, provided that the 
%resulting surface brightness falls below the very faint limits
%set by that study. Indeed, point sources with colors consistent with
%those of blue HB stars can be traced beyond the limiting radii
%measured by Odenkirchen et al.\ (2001a; see also Section
%\ref{subsec:morphology-draco}), however, their number densities are 
%so low that it is unclear whether we are tracing Draco members or
%field stars here.

We choose the SDSS color $( g^{\star} - i^{\star})$ to ensure a 
large color baseline in two of the most sensitive SDSS filters.  
(Note that the superscript $^{\star}$ is conventionally used to
indicate that the
photometric calibration of the SDSS EDR data is not yet final and
may be further refined in later data releases; see also Stoughton
et al.\ 2002.)  The SDSS g filter lies between Johnson B and 
V and largely overlaps with these two filters, while the SDSS i
filter is slightly more narrow and a bit bluer than the Kron-Cousins I
filter (see Fig.\ 1 in Grebel 2002).  Transformation
equations between the SDSS filter system and the conventional 
Johnson-Cousins system are given in Fukugita et al.\ (1996); for the 
purposes of our study we will consider SDSS $g^{\star}$ as
equivalent to Johnson $V$.    

All SDSS magnitudes were corrected for Galactic foreground extinction
using the appropriate values of Schlegel et al.\ (1998), which are provided by
the SDSS database.  We show the distribution of Galactic reddening 
in the region of Draco and surroundings in Fig.\ 1 of Odenkirchen et al.\ 
(2001a).  We assume that there is no internal extinction in Draco
(see, e.g., Gallagher et al.\ 2003 and references therein).  The resulting 
dereddened 
SDSS magnitudes are denoted $g^{\star}_0$ and $i^{\star}_0$.
To select blue HB stars, we considered only objects
flagged as stars in the database, which in addition were required to
fulfill the criteria

$$19\fm8 < g^{\star}_0 < 20\fm3 \,\,\,\, {\rm and} \,\,\,\, -0\fm35 < (g^{\star}_0 - 
i^{\star}_0) < -0\fm10$$
$$19\fm8 < g^{\star}_0 < 20\fm7 \,\,\,\, {\rm and} \,\,\,\, -0\fm50 < (g^{\star}_0 - 
i^{\star}_0) < -0\fm35$$ 

for the ``red'' part of the blue HB and for the extended
blue HB, respectively (Fig.\ \ref{Draco_sample_CMD},
fat dots).  We note that while extinction properties vary with stellar
temperature, these effects are anticipated to be small within the range
of colors considered here and correspond to an uncertainty of at most 
0.01 mag in $g^{\star}_0$ (Grebel \& Roberts 1995, their Figure 6, top
left panel; assuming that $(M-T_2) \sim (g^{\star}_0 - i^{\star}_0)$
and $g^{\star}_0 \sim M$).

\begin{deluxetable}{cccccc}
\tabletypesize{\scriptsize}
\tablecaption{Blue horizontal branch width of Draco.
\label{Tab_width}}
\tablewidth{0pt}
\tablehead{
\colhead{Area [$^{\circ} \times ^{\circ}$]} & 
\colhead{$\langle \Delta g^{\star}_0 \rangle _{\rm HB}$ [mag]}   
& \colhead{Number of bins}   &
\colhead{Number of stars} &
\colhead{predicted $\langle \Delta V \rangle _{\rm HB}$ [mag]}\tablenotemark{a} &
\colhead{acceptance level\tablenotemark{b}}
}
\startdata
$0.25 \times 0.25$ & 0.12 & 5 & 17 & $0.125\pm0.023$ & 48.0 \\
$0.50 \times 0.50$ & 0.13 & 8 & 35 & $0.138\pm0.023$ & 53.8 \\
$1.00 \times 1.00$ & 0.13 & 8 & 48 & $0.162\pm0.020$ & 81.8 \\
$1.50 \times 1.50$ & 0.14 & 8 & 56 & $0.185\pm0.018$ & 91.5 \\
$2.00 \times 2.00$ & 0.13 & 8 & 60 & $0.207\pm0.017$ & 99.3 \\
$2.50 \times 2.50$ & 0.14 & 8 & 63 & $0.225\pm0.016$ & 99.5 \\
\enddata
\tablenotetext{a}{Predicted average HB width $\langle \Delta V \rangle
  _{\rm HB}$  for the  best-fit tidal model {\em 4} obtained from
  bootstrapping.} 
\tablenotetext{b}{Percentage of bootstrapping attempts that delivered
  $\langle \Delta V \rangle _{\rm HB}$ in excess of the observed value
  in column 2.} 
\end{deluxetable}
We then applied the same procedure as used in Klessen \& Zhao (2002) 
to calculate the luminosity width of the HB:  We 
calculated the average over the standard deviations $\sigma(g^{\star}_0)$
of stars in $(g^{\star}_0 - i^{\star}_0)$ color bins of a width of
$\Delta (g^{\star}_0 - i^{\star}_0) = 0.05$ mag for all blue HB 
star candidates within certain distances from the center of Draco.
The results are given in Table \ref{Tab_width}.  Column 1 shows the 
selection area centered on Draco, column 2 the average of the standard
deviations in $g^{\star}_0$, column 3 specifies the number of 
$(g^{\star}_0 - i^{\star}_0)$ color bins, and column 4 the number of
blue HB candidates that entered into the calculations. For comparison, column
5 gives the HB thickness, $\sigma(V)$, expected from the best-fit tidal model
(see Section \ref{subsec:HB-thickness}).
Table \ref{Tab_width} demonstrates that the blue HB exhibits only a 
small width in the center of Draco
and that it remains narrow also when larger areas are
considered, while tidal models predict a significant HB widening.  The fractional increase in blue HB star candidates 
in the outermost annulus (see column 4) is consistent with the numbers
expected from field contamination quoted above.

\section{A Tidal Model for the Draco?}
\label{sec:model}

With regard to its morphology and structure,
Draco is one of the best studied dSph satellite galaxies
of the Milky Way.  Its compact
and seemingly undistorted morphological appearance, together with a
well-defined, thin HB in the CMDs sets
severe constraints on tidal models. In this Section we investigate
whether it is possible to construct a tidal model for Draco that is
able to reproduce {\em all} observed morphological, kinematic, and
photometric properties of this satellite galaxy.
\begin{deluxetable}{crcc}
\tablecaption{Initial properties of the model satellite galaxies.
\label{Tab_properties}}
\tablewidth{0pt}
\tablehead{\colhead{Model} & \colhead{$r_{\rm c}$ [kpc]} &
\colhead{$r_{\rm t}$ [kpc]} & \colhead{M [$M_{\odot}$]} }
\startdata
{\em 1}    & 0.3   & 1.5   & $10^7$\\
{\em 2}    & 0.1   & 0.5   & $10^6$\\
{\em 3}    & 0.05  & 0.5   & $10^6$\\
{\em 4}    & 0.02  & 0.2   & $10^6$\\
\enddata
\end{deluxetable}

\subsection{The Numerical Method}
\label{subsec:numerics}
\begin{figure}
\plotone{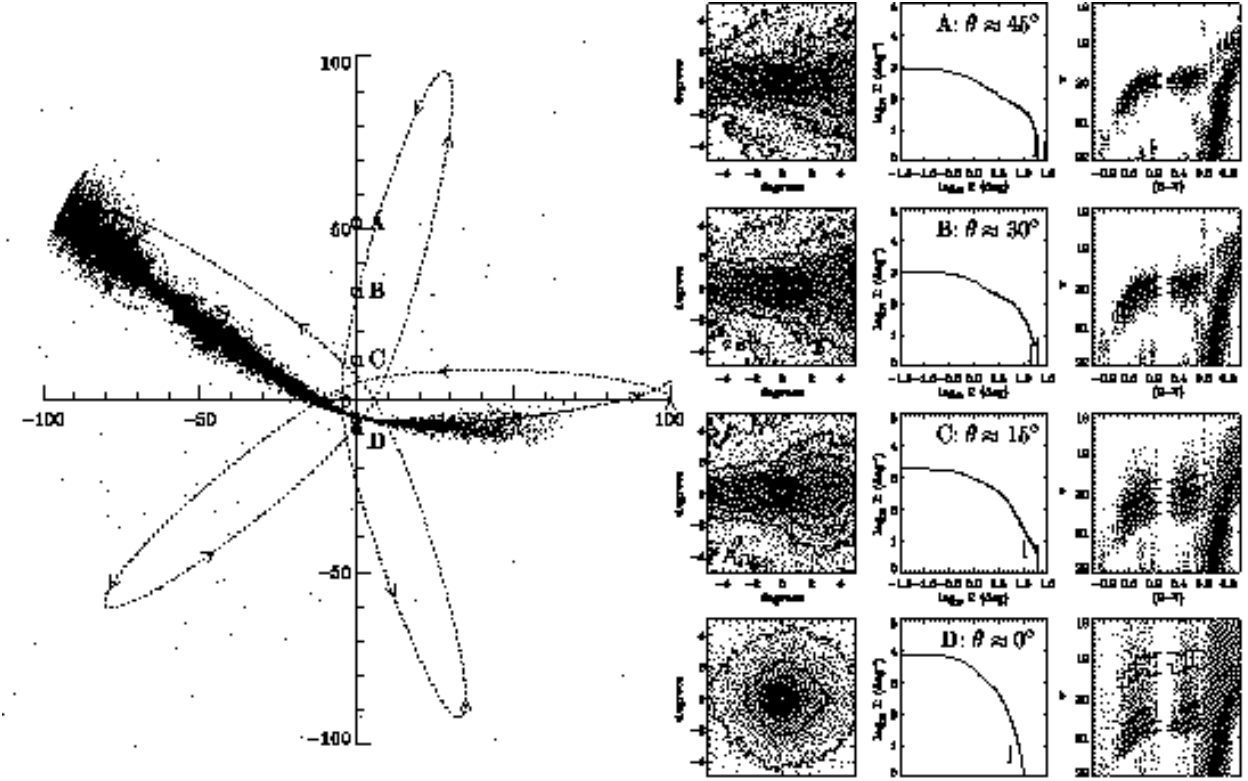}
\caption{\label{fig:trajectory}\small  The effect of projection on
  morphological appearance. The left side shows the distribution of
  stars in Model {\em 1} in the tidal field of the Milky Way seen at
  $t=4.5\,$Gyr, i.e., about $2\,$Gyr after the complete disruption of
  the satellite galaxy. The galaxy has dissolved into extended tidal
  tails on each side of the maximum of the stellar density tracing the
  location of the former core of the satellite (as indicated by the
  dotted circle). The trajectory of the satellite is indicated by the
  dotted line where the arrows indicate its location in intervals of
  $\Delta t=0.5\,$Gyr. The orbital eccentricity is $\epsilon =0.9$,
  and the satellite starts at $t=0\,$Gyr at the right at a
  galactocentric distance $d_{\rm gal}=100\,$kpc with velocity
  $v=25\,$km$\,$s$^{-1}$ in azimuthal direction. The axes indicate a
  galactocentric coordinate system with units in kpc. The location of
  the hypothetic terrestrial observer at $d_{\rm gal}=8.5\,$kpc is
  indicated by the black dot on the vertical axis. The viewing angle
  $\theta$ under with the tidal satellite is seen from this position
  is almost zero.  This leads to very compact morphological appearance
  of the system when projected onto the observers plane of the sky and
  a large depth along the line of sight leading to considerable
  widening of the distribution of stars on the horizontal branch in
  the color-magnitude diagram. Three other positions for observing the
  satellite are indicated by open dots along the vertical axis,
  corresponding to $\theta \approx 15^{\circ}$, $30^{\circ}$, and
  $45^{\circ}$. The right hand side of the figure illustrates how the
  tidal debris appears on the sky of the hypothetical observer for the
  four projection angles. The first column depicts the surface density
  distribution of particles. The contour lines are linearly spaced in
  intervals of 10\% of the central value. The dashed contour indicates
  the 1\% level. The second column shows the azimuthally averaged
  radial number density profile $\Sigma$. The arrow gives the distance
  at which $\Sigma$ has dropped by two orders of magnitude. The third
  column depicts the resulting distribution of stars in the
  color-magnitude diagram (using the globular cluster M3 as template,
  see Klessen \& Zhao 2002). The lines indicate the resulting
  horizontal branch thickness as determined from the full
  $5^{\circ}\times5^{\circ}$ field (outer pair) and the central
  $0.5^{\circ}\times0.5^{\circ}$ (inner pair), see Section
  \ref{subsec:HB-thickness} for further details.  Note that model {\em
    1} is selected for illustration purposes, it is not an adequate
  model for Draco.  }
\end{figure}

In order to construct and constrain a tidal model for the Draco dSph
galaxy, we follow the evolution and disintegration of various
satellite dwarf galaxies in the tidal field of our Galaxy until well
after they have completely dissolved. The considered model galaxies vary in
mass and size, but all follow an initial Plummer-type equilibrium
density profile with parameters as summarized in Table
\ref{Tab_properties}. The satellites are placed on nearly radial
orbits with eccentricity $\epsilon =0.9$ and peri- and apogalactic
distances $100\,$kpc and $5\,$kpc, respectively (see Figure
\ref{fig:trajectory}).  The model galaxies have a fixed mass-to-light
ratio $M/L = 3$ and there is {\em no} dark matter component.  The
Milky Way, however, does contain dark matter. It dominates the
gravitational potential at large scales, and is described as
isothermal sphere with circular velocity of $220\,$km$\,$s$^{-1}$ and
a core radius of 5$\,$kpc.

Each satellite galaxy consists of 131$\,$072 particles, and their time
evolution in the Galactic tidal field is followed numerically solving
the equations of motion using a standard TREECODE scheme (Barnes \&
Hut 1986). The Galactic tidal field hereby is treated as rigid
external potential. Further details are given
in Klessen \& Kroupa (1998).

\subsection{The Effect of Projection on Morphology}
\label{subsec:projection}
The first constraints on tidal models for Draco can be derived from
the very compact observed morphological appearance of the dwarf galaxy
and the lack of `extra-tidal' stars. The radial profile of stellar
density of Draco is well described by King models (King 1962, 1966) or
generalized exponentials (e.g., S\'ersic 1968) with core radii of
$r_{\rm c}\approx 8'$ and limiting radii $r_{\rm t}$ in the range $40'$ to
$50'$ (Odenkirchen et al.\ 2001a; Aparicio et al.\ 2001). Adopting a
distance to Draco of $d=80\,$kpc, this corresponds to a total radius
of the galaxy of about $1\,$kpc. Note that the available
observational data do not imply that a clear outer `edge' in the galaxy has
been detected (Section \ref{subsec:morphology-draco}). 
The only firm statement that can be made is that the
stellar density drops to less than 1\% of its central value within
$r_{\rm t}\sils 50'$, where the galaxy blends into the distribution of
foreground stars.

The angular size of Draco is very small compared to typical values
reported from numerical simulations. The tidal models published to
date predict limiting radii considerably in excess of $1^\circ$, when
the satellite galaxy is projected onto the plane of the sky of an
hypothetical observer on Earth after its complete dissolution, i.e., in
the phase when it clearly bears morphological and kinematic
resemblance to Galactic dSphs (e.g., Figure 2 of Klessen \& Zhao
2002; see also Kroupa 1997, Klessen \& Kroupa 1998). Furthermore,
these model dSphs often appear quite elongated along the orbital
trajectory and have no clear truncation radius, i.e., exhibit a large
number of `extra-tidal' stars.
The fact that Draco is compact (i.e., the surface density decreases by
two orders of magnitude within $r_{\rm t} \sils 50'$) and not very
elongated (with a ratio between minor and major axis of $\sim 0.7$)
therefore strongly restricts the overall parameter space for tidal
models. 

Let us first consider the effect of the orbital parameters of tidal
satellite galaxies on their morphological appearance for observers on
Earth. As a dwarf galaxy orbits around the Milky Way, it loses
stars, which subsequently populate two extended tidal tails. After complete
disruption the location of the former satellite is still discernible
for one to two Gyr as an enhancement of stellar density, moving along the
orbital trajectory in between the leading and trailing tidal arm. This
remnant slowly expands and decreases in density as its stars 
drift apart and populate the tidal arms. If the satellite is on a relatively
circular orbit, a terrestrial observer (being much closer to the
Galactic center) would see the resulting elongated structure of the
tidal debris from the `side', and would recognize a low-density feature that
stretches over many degrees on the plane of the
sky  along the orbital path, bearing little resemblance of real dSph galaxies. In
addition, it would be difficult to see and distinguish this feature at
all, as it is likely to blend into the distribution of Galactic
foreground stars. In this case only an extensive search for kinematic
signatures of tidal streamers can reveal its true nature (see,
e.g., Majewski et al.\ 2000; Helmi \& de~Zeeuw 2000; Dohm-Palmer et
al.\ 2001; Helmi, White, \& Springel 2002; Kundu et al.\ 2000; Newberg
et al.\ 2002).
 On the other hand, if the satellite
was originally on a very eccentric orbit (as the model galaxies
discussed here), it is likely that the viewing angle between the
orbital trajectory and the line of sight of an observer on Earth is
very small. The tidal debris is then seen along the major axis and
subsequently covers a much smaller projected area on the observer's plane
of the sky, with morphological and kinematic features now similar to
observed dSph galaxies (Kroupa 1997; Klessen \& Kroupa 1998).
This `piling up' of
stars along the line of sight makes the tidal remnant easier to detect
against the Galactic foreground stars, and leads to a large dispersion
of stellar distance moduli, which in turn results in a widening of the
distribution of stars in the CMD along the
luminosity axis. This widening can be particularly well
measured using HB stars, and has been suggested
as a possible observational test for the tidal models (Klessen \& Zhao
2002).

Figure \ref{fig:trajectory} illustrates the effect of the orbital 
eccentricity $\epsilon$ on the projected morphological appearance. Instead of
observing satellite galaxies on different orbits (from the same
position at $d_{\rm gal} = 8.5\,$kpc), we equivalently select a model
with $\epsilon = 0.9$ and vary the location of the observer, such that
the line of sight covers a range of viewing angles $0^{\circ} \sils
\theta \le 45^{\circ}$ with the orbital trajectory (and hence the
major axis of the tidal debris). The right-hand side of Figure
\ref{fig:trajectory} shows that only for eccentricities that allow for
$\theta \sils 5^{\circ}$  tidal models are able to reproduce a compact
and roundish morphological appearance with azimuthally averaged
radial stellar density profiles.  These profiles
can be fit by King models or generalized
exponential models and exhibit a lack of significant numbers of
`extra-tidal' stars.  This limits the parameter space for tidal models
of Draco to high orbital eccentricities. If Draco currently were near its
perigalacticon as indeed suggested by Kleyna et al.\ (2001), then tidal models
could immediately be excluded.
Note that small viewing
angles favor projections where the major axis is perpendicular to the
orbital trajectory, leading to line of sight velocity gradients that
run primarily along the minor axis (although connected to very large
dispersions, see Section \ref{subsec:vLOS}). This is in contrast to angles
$\theta > 10^{\circ}$ where the elongation always lies along the
orbital path (similar to the line of sight velocity gradient).  
%{\bf \{} Note also that for $\theta \sils 5^{\circ}$ the extent along the
% line of sight is largest and the widening of the stellar distribution
% in the color magnitude diagram is most significant. {\bf \}}
\begin{figure}
\plotone{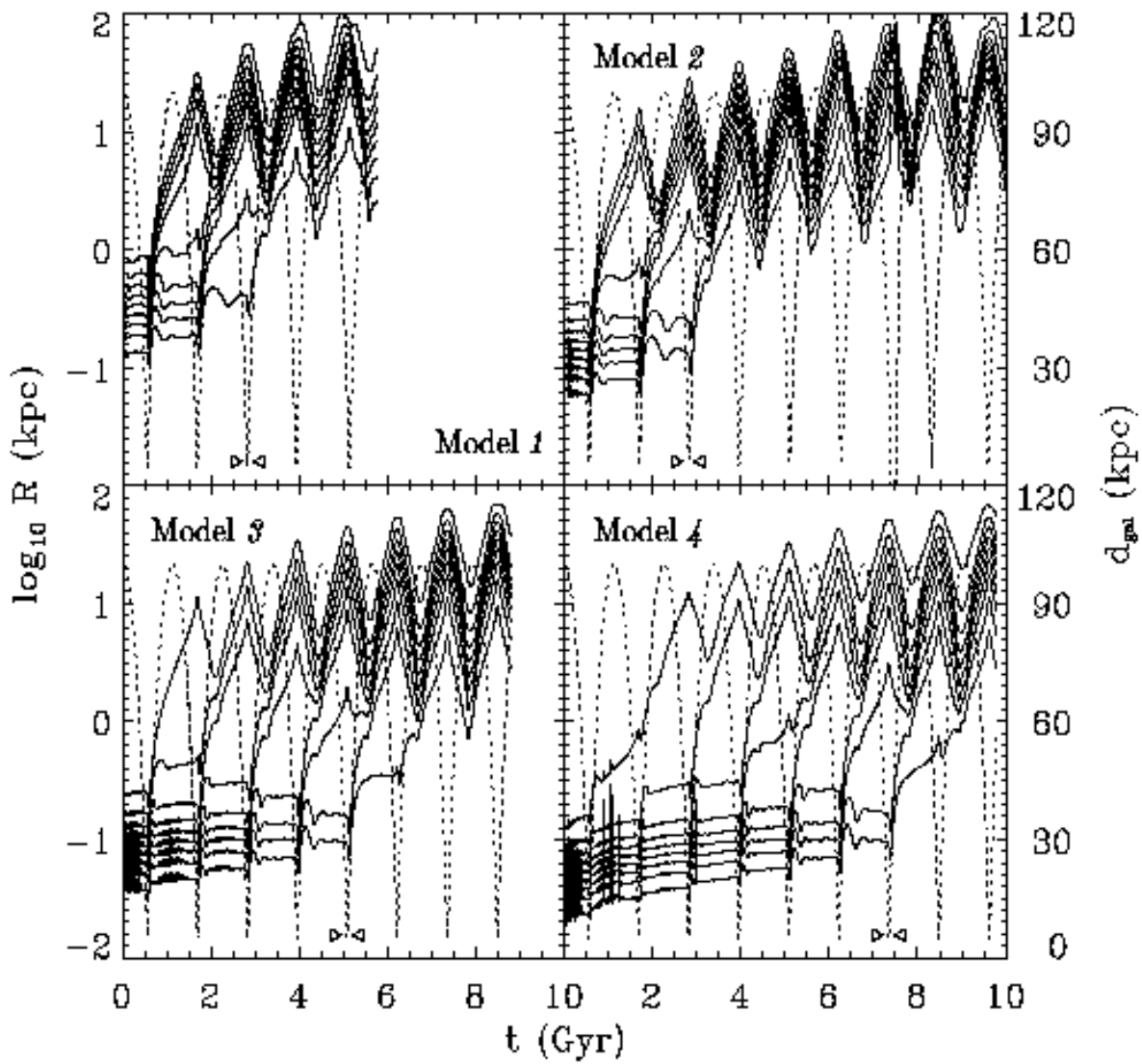}
\caption{\label{fig:Lagrangian}\small Lagrangian radii as function of
time. The figure shows the radii (left axis) containing 10\%, 20\%,
\dots 90\% of the mass as the satellite galaxy dissolves in the tidal
field of the Milky Way. The corresponding galactocentric distances are
indicated by dotted lines (right axis), and the perigalactic passage
that leads to complete dissolution is indicated by triangles.}
\end{figure}

\subsection{Small Angular Size and the Timescale for Tidal Disintegration} 
\label{subsec:ang-size}
We selected model {\em 1} for illustration in Figure
\ref{fig:trajectory}, because it is the largest satellite in our suite
of models and it is the one most easily disrupted. Note that 
already in the initial bound state, it is too extended to describe Draco at
a distance of $80\,$kpc. It is therefore necessary to study the tidal
evolution and observational appearance of satellite galaxies of
smaller sizes, to see whether it is possible to find a model that
better resembles the observational data. Galaxies {\em 2} to {\em 4}
are less massive than model {\em 1} by a factor of 10
($10^6\,$M$_{\odot}$ instead of $10^7\,$M$_{\odot}$) and are
increasingly more compact ranging in their initial stage in size from
$r_{\rm t}=0.5\,$kpc to $r_{\rm t}=0.2\,$kpc with core radii from
$r_{\rm c}=0.1\,$kpc down to $r_{\rm c}=0.02\,$kpc, respectively (see
Table \ref{Tab_properties}). Model {\em 4} is therefore quite
comparable to a massive globular cluster (Harris 1996). The time 
evolution of the Lagrangian radii encompassing
10\%, 20\%, etc., of the mass is shown in Figure \ref{fig:Lagrangian}
as the satellite galaxies orbit around the Milky Way. Models {\em 1}
and {\em 2} dissolve after $t \approx 2.8\,$Gyr during their third
perigalactic passage. Model {\em 3} needs five ($t \approx 5.2\,$Gyr)
and model {\em 4} seven passages ($t \approx 7.5\,$Gyr) to completely
dissolve in the Galactic tidal field. Satellite galaxies of the same
mass that are initially more compact would not fully dissolve over the
lifetime of the Galaxy on the orbit considered here. Similar characteristics
hold for
satellites that are more massive but have a comparable size as model
{\em 4}. This limits the available parameter space of initial
satellite properties ($M$, $r_{\rm c}$, and $r_{\rm t}$) and motivates
our choices in Table \ref{Tab_properties}.

%\vspace*{0.5cm}
\begin{figure}[ht]
\plotone{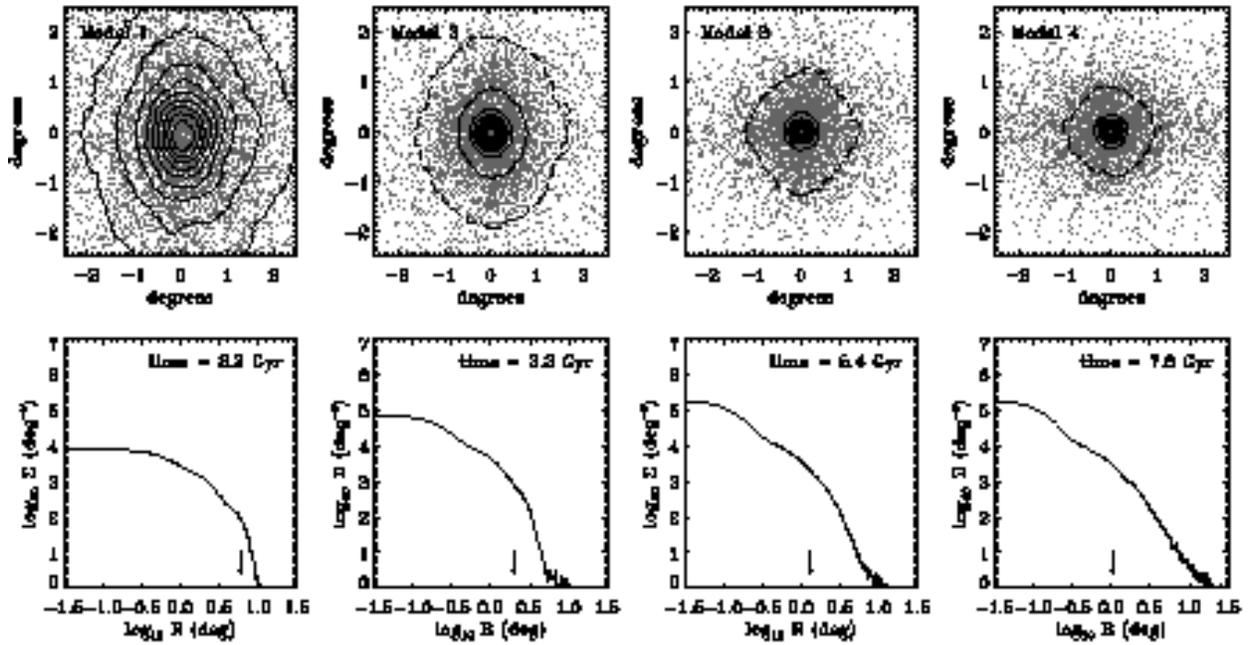}
\caption{\label{fig:compactness}\small Images of the model galaxies
shortly after
complete disruption. The upper panel shows the surface density of the
model galaxies  shortly after their perigalactic passage that
fully disrupts them, at a time when they reach again a distance to the
observer of $d\approx80\,$kpc. Continuous contours are overlayed onto
the projected distribution of $N$-body particles (gray dots). The
contour spacing is linear in  steps of 10\% of the central value. The dashed line
indicates the 1\% level.  The lower panel depicts the corresponding
azimuthally averaged radial profile. $\Sigma$ is defined as the number
density of $N$-body particles per square degrees. The conversion into
stellar surface density is discussed in the text. }
\end{figure}

Within the allowed parameter range smaller initial
satellites lead to more compact morphological appearance of the tidal
remnant after complete dissolution. In our set of four satellite
galaxies, only model {\em 4} is able to reproduce the fact that the
stellar surface density of Draco falls off to less than 1\% within
$\sils 1^{\circ}$. Figure \ref{fig:compactness} illustrates this point
by showing the image of the satellites as seen on Earth after complete
disruption at the time when they first reach distances to the observer
of $d\approx 80\,$kpc. The upper panel shows the projected number
density $\Sigma$ of $N$-body particles per square degree on the sky;
the lower panel shows the resulting azimuthally averaged radial
profile. The conversion of $\Sigma$ into the corresponding {\em
  stellar} number density $\Sigma_*$ depends on the number of
particles used in the simulation, the mass of the galaxy, and
assumptions about the IMF. A prediction of the observed surface
density of stars (e.g., for comparing with star count results)
furthermore depends on the sensitivity limit of the instrument used
and the crowding (number density) of foreground stars, which gives a
constant offset.  For instance, the number density of particles in the
central region of model {\em 4} is $\Sigma \approx 1.7\times 10^5$ per
square degree.  As the number of particles is $131\,072$ and the total
mass $10^6\,$M$_{\odot}$, each particle carries $\sim
7.6\,$M$_{\odot}$.  We assume a standard IMF as proposed by Kroupa
(2002) with average stellar mass $\langle M \rangle =
0.38\,$M$_{\odot}$ and where stars with $ M \ge 1\,$M$_{\odot}$
contribute roughly 50\% of the mass but only 6\% in number. If we
furthermore assume a detection limit of $1\,$M$_{\odot}$, then each
$N$-body particle represents about $1.2$ stars 
% on the main sequence
lying above the detection threshold, $\Sigma_*= 1.2 \Sigma$. For the
very inner region of Draco, model {\em 4} thus predicts star counts of
order $10^5$ per square degree. This is consistent with the observed
values (Irwin \& Hatzidimitriou 1995; Odenkirchen et al.\ 2001a;
Aparicio et al.\ 2001).

\begin{figure}
\plotone{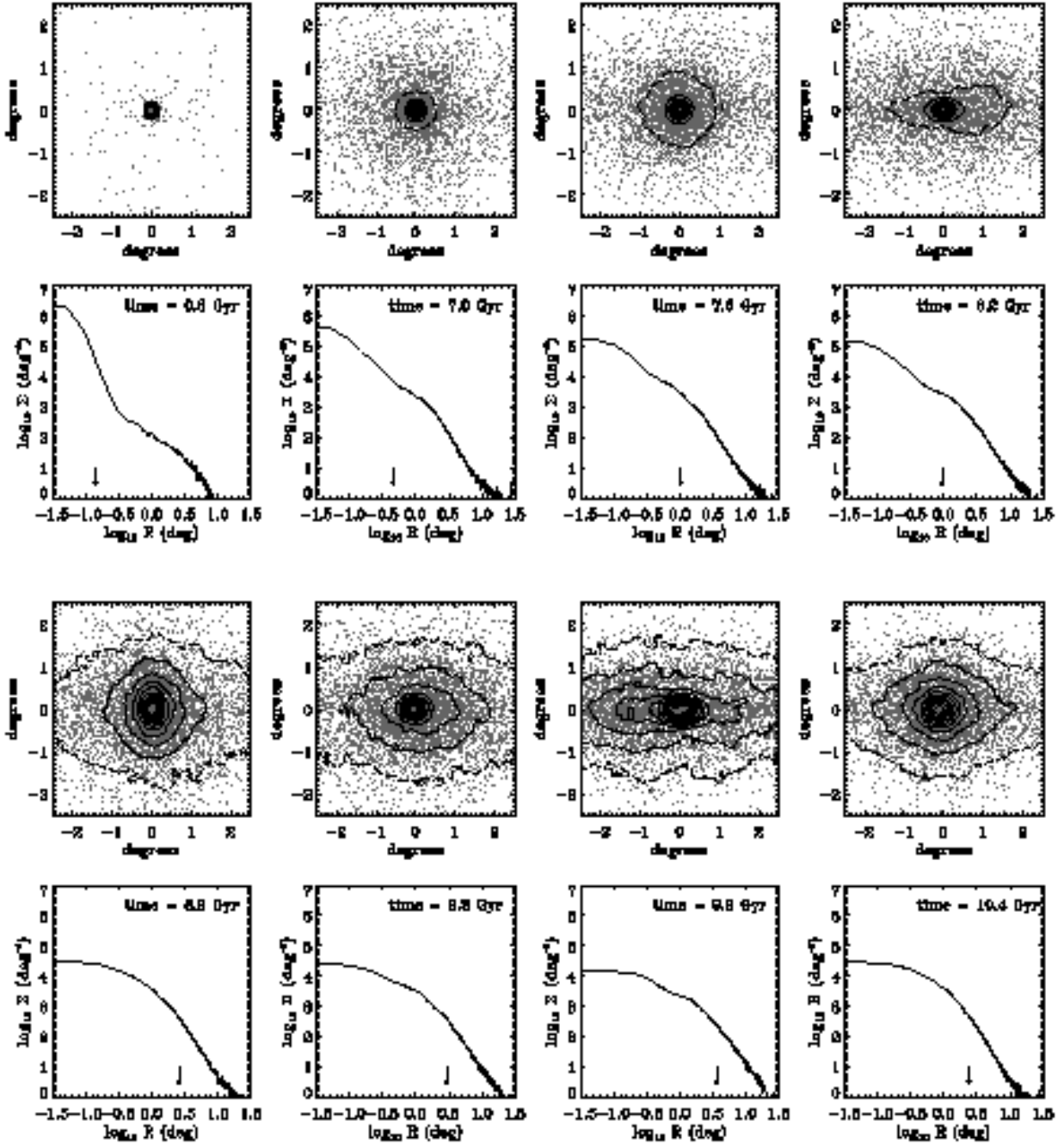}
\caption{\label{fig:time-evolution}\small Morphological appearance of model
  {\em 4} as function of time. The figure describes the projected particle
  distribution (upper panels) and radial surface density profile (lower
  panels) at times when the distance between observer and the density center
  of the satellite galaxy reaches $d\approx80\,$kpc. All such instances are
  shown after the satellite completely dissolves in the perigalactic passage
  at $t\approx 7.5\,$Gyr. At earlier times, the satellite
  exhibits no substantial signs
  of evolution in projection and we show only two snapshots, one
  immediately
  after the start of the calculation, $t=0.3$, and the other
  before complete disruption, 
  $t=7.0$. }
\end{figure}

\begin{figure}
\plotone{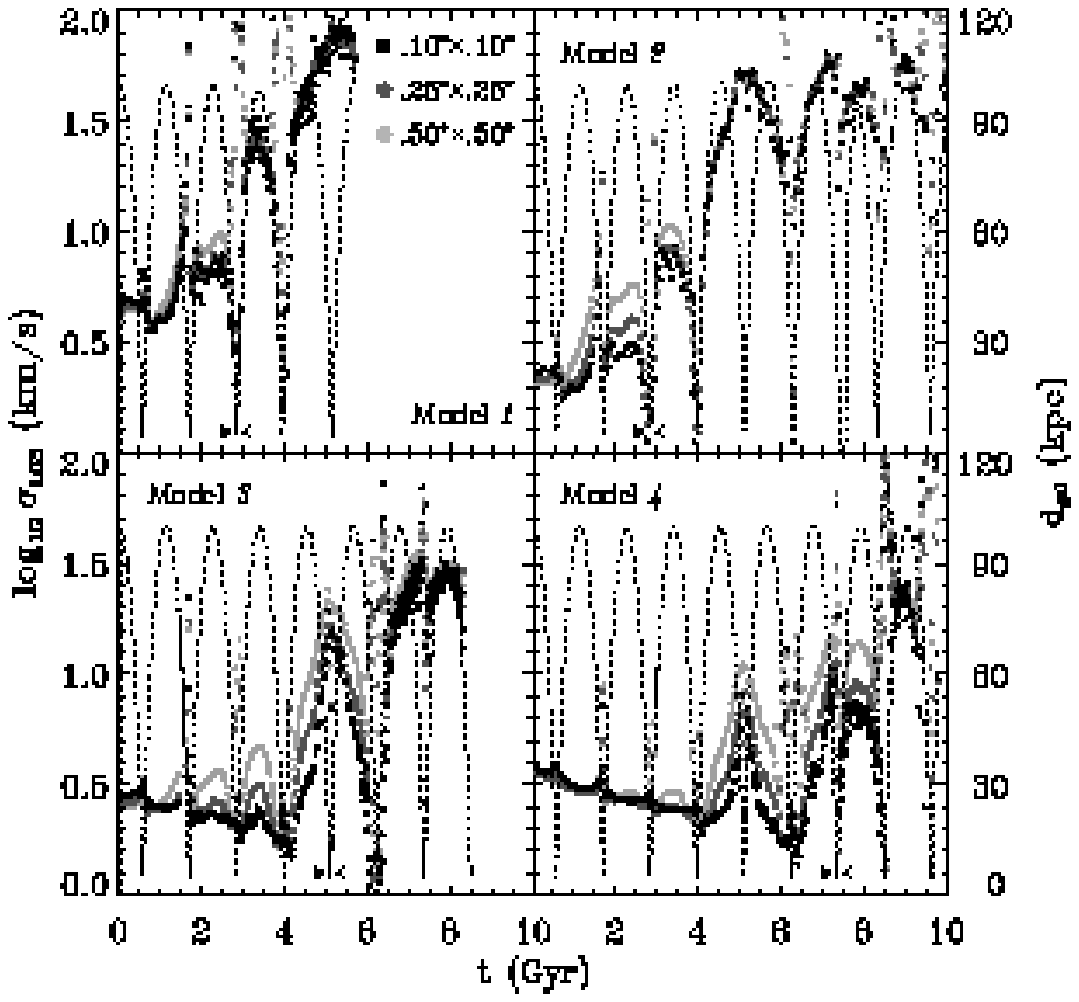}
\caption{\label{fig:vLOS}\small  Time evolution of the line of sight
velocity dispersion $\sigma_{\rm LOS}$. For all four model galaxies,
$\sigma_{\rm LOS}$ vs.\ $t$ is calculated for three different
observational field sizes (as indicated in gray, units on the left
axis). Before complete tidal disruption, the measured velocity
dispersion is lowest when only considering a small central field, as
$\sigma_{\rm LOS}$ is dominated by the stars in remaining bound 
core. With increasing field size also unbound stars in the tidal tails
begin to contribute and $\sigma_{\rm LOS}$ increases. At about
$1\,$Gyr after disruption as the system becomes sufficiently dispersed,
the influence of the central remnant vanishes and the variation of
$\sigma_{\rm LOS}$ obtained from different field sizes becomes negligible. To
see the relation between the observable $\sigma_{\rm LOS}$ and the
galactocentric distance, $d_{\rm gal}$ is indicated by the dotted
lines (right axis).  The
perigalactic passage that leads to complete dissolution is indicated
by triangles. }
\end{figure}

\subsection{Rapid Expansion after Complete Tidal Dissolution}
\label{subsec:vLOS}
The results of the previous section confirm that it is possible to
construct tidal models that resemble the morphology of the Draco dSph
galaxy at least immediately after their complete disruption. However,
how long does this period of time last?  This question is addressed in
Figure \ref{fig:time-evolution} which shows the projected surface
density of model {\em 4} during various phases of its dynamical
evolution under the condition that the galaxy lies at a distance of roughly
$80\,$kpc to an observer on Earth.  The figure depicts all such
instances in the time interval $t>5\,$Gyr until the end of the
simulation at $t=11.5\,$Gyr. Earlier times exhibit no substantial
evolution in the projected stellar density distribution and only two
snapshots are shown. Note that the satellite completely dissolves
during the seventh perigalactic passage at $t\approx 7.5\,$Gyr. The
figure illustrates how the tightly bound satellite galaxy evolves into
unbound debris in the tidal field of the Galaxy, and how this tidal
stripping process influences the observable structure. Note the
variation of morphological parameters (e.g., in the axis ratios, the
position of the major axis, or the `smoothness' of the particle
distribution) amongst the six snapshots with $t>7.5\,$Gyr. Note also
that only the first two snapshots after complete disintegration are
compact enough to describe Draco. At later times the surface density
does not drop below 1\% of its central value within radii of $\sim
2.5^{\circ}$, largely exceeding the limiting radius of Draco $r_{\rm
t}\sils 50'$. Figure \ref{fig:time-evolution} therefore indicates that
the time interval during which tidal models are able to reproduce
properties of Draco is quite limited (to about $1\,$Gyr). This implies
that we would observe the dSph galaxy at a very special instance in
time (namely right after its tidal disruption) if the tidal models were
correct.  This result is further supported by the time evolution of
the line of sight velocity dispersion $\sigma_{\rm LOS}$, as indicated
in Figure \ref{fig:vLOS}.  For model {\em 4}, $\sigma_{\rm LOS}$ lies
within the observed margin $8\,$km$\,$s$^{-1} < \sigma_{\rm LOS} \sils
10\,$km$\,$s$^{-1}$ only for a period of about 1 Gyr following the
complete disruption. At earlier times $\sigma_{\rm LOS}$ is smaller,
and at later times it is larger. Similar time scales hold for the other 
models.

\subsection{Horizontal Branch Thickness}
\label{subsec:HB-thickness}
\begin{figure}[t]
\plotone{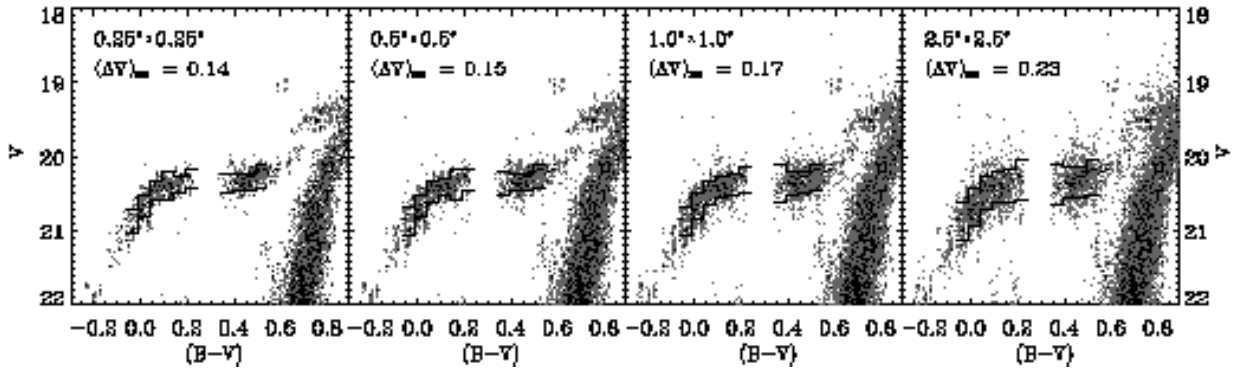}
\caption{\label{fig:HB-4-fields}\small Influence of the field size
on the horizontal branch thickness. The figure shows the color-magnitude
diagram of model {\em 4} at $t=7.6\,$Gyr using the globular
cluster M3 as template for field sizes ranging from
$0.25^{\circ}\times0.25^{\circ}$ to $2.5^{\circ}\times2.5^{\circ}$
centered on the maximum of the surface density of stars. The
horizontal branch thickness $\langle \Delta V \rangle_{\rm HB}$ is
defined as the average value of 
the standard deviation of $V$
for horizontal branch stars determined from bins of width
$\Delta(B-V)=0.05$ mag in the range $-0.15$ mag $\le (B-V) \le 0.5$ mag (see
Section \ref{subsec:HB-draco}).  }
\end{figure}

As discussed above, the compactness of Draco at a distance of
$80\,$kpc, the shape of its radial surface density distribution, and
the measured value of its line-of-sight velocity dispersion severely
limit the available parameter space for tidal models and make their
applicability to Draco very unlikely. To fully exclude a tidal model
for Draco, one needs to consider its HB. As discussed in Section
\ref{subsec:HB-draco}, the HB of stars in the CMD is remarkably thin
with $\langle \Delta g^{\star}_0 \rangle_{\rm HB} \approx 0.13\,$mag,
and it shows remarkably little variation with extent of the region
considered. Values range from $\langle \Delta g^{\star}_0 \rangle_{\rm
  HB} \approx 0.12\,$mag for stars within a radius of $0.25^{\circ}$
around the center of Draco to $\langle \Delta g^{\star}_0 \rangle_{\rm
  HB} \sils 0.14\,$mag when considering all HB stars within
$2.5^{\circ} \times 2.5^{\circ}$.  The best-fit tidal model exhibits
comparable values only in the innermost region. When considering
larger areas, it shows a HB thickness significantly larger than
observed.  It is a general property of tidal models to predict a
strong correlation between the HB thickness and the angular size of
the considered field. This is in sharp contrast to the observational
data of Draco which reveal no significant variation of HB thickness
with area.  As illustration, Figure \ref{fig:HB-4-fields} shows the
CMD obtained for fields of different sizes from model {\em 4} at time
$t=7.6\,$Gyr when it reaches a distance of $80\,$kpc for the first
time after complete disruption. The corresponding stellar density
distribution can be seen in the right column of Figure
\ref{fig:compactness}.  The synthetic CMDs are obtained using the
method described by Klessen \& Zhao (2002) with the globular cluster
M3 as template, and with $\langle \Delta V \rangle_{\rm HB}$ computed as
in Section \ref{subsec:HB-draco}.  M3 has metallicities similar to
Draco and has a very thin HB with $\langle \Delta V \rangle_{\rm HB}
\sils 0.08\,$mag.  Hence, any significant HB widening apparent in
Figure \ref{fig:HB-4-fields} stems from stellar distance moduli
variations in the projected tidal debris.  The larger the considered
area, the larger is the derived HB thickness.  Stars that are further
away from the projected center of density are likely to be further
downstream or upstream along the tidal arms. The inclusion of those
stars into the analysis leads to a larger range of distance moduli and
subsequently to a widening of the luminosity distribution in the CMD.

\begin{figure}
\plotone{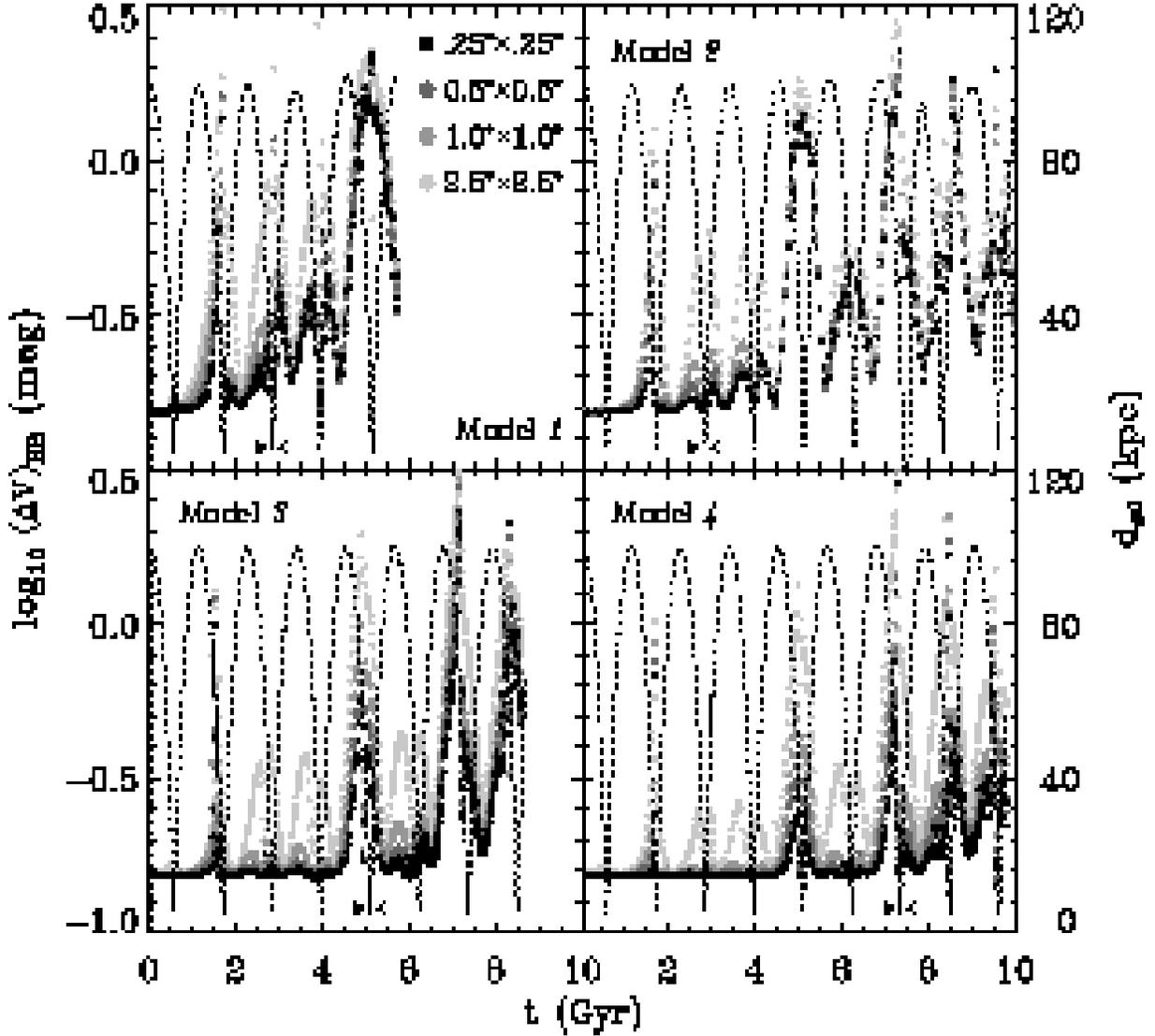}
\caption{\label{fig:HB}\small Horizontal branch thickness as function of
time. For models {\em 1} to {\em 4} the figure shows $\langle \Delta
V \rangle_{\rm HB}$ vs.\ $t$ calculated for four different field sizes
(as indicated in gray, left axis). The galactocentric distance $d_{\rm
gal}$ at the corresponding times is indicated by the dotted lines
(right axis). As already indicated in Figure \ref{fig:HB-4-fields},
larger field sizes lead to larger horizontal branch thickness. The
perigalactic passage that leads to complete dissolution is indicated
by triangles.
}
\end{figure}
To test the statistical significance of our result we need to address
two questions. First, how does the HB thickness of the best-fit tidal
model vary within the evolutionary window allowed from morphological
and kinematic considerations? Second, what is the effect of the
small sampling size in the observations -- the HB thickness for the
innermost region of Draco is obtained from only 17 BHB stars? In the
previous section we have demonstrated that the LOS velocity dispersion
in the best-fit tidal model lies within the observed margins only for
a period of about $1\,$Gyr after complete disruption. Furthermore,
Draco is at a distance of $\sim80\,$kpc. Inspection of the satellite
trajectories in Figure \ref{fig:Lagrangian} reveals that model {\em 4}
complies with both requirements only twice, at $t=7.6\,$Gyr and at
$t=8.2\,$Gyr. When we allow for a distance uncertainty of Draco of
5\%, which corresponds in the models to a time span of roughly
$50\,$Myr, and compare the HB morphology within and between these two
intervals, we see that the variations are very small. This enables us
to derive a well-defined and meaningful average HB thickness.  The
second question can be addressed by bootstrapping. From the large
number of HB stars in the model we randomly select a small subset
corresponding in number to the observed sample (e.g.\ 17 for the
innermost $0.25^{\circ}\times0.25^{\circ}$) and compute the resulting
HB width. We repeat the procedure $10\,000$ times for each field size.
The resulting HB thickness $\langle \Delta V \rangle_{\rm HB}$
together with its standard deviation averaged over all bootstrapping
attempts and averaged over all allowed snapshots of model {\em 4} is
listed in column 5 of Table \ref{Tab_width} for direct comparison with
the observations. The statistical error decreases slightly with
increasing field size because of the increasing number of stars
considered in the subsamples (column 4).  The last column in Table
\ref{Tab_width} indicates in addition the percentage of the
bootstrapping attempts that delivered $\langle \Delta V \rangle_{\rm
  HB}$ values in excess of the observations. This number increases
from $\sim 50$\% in the inner $0.25^{\circ}\times0.25^{\circ}$ to
about $100$\% in the full field, and gives a clear indication that we
must consider wide-field data to arrive at statistically sound
statements about the nature of dSph galaxies. It should be noted, that
the results still hold, even if no assumption about the intrinsic HB
morphology are made, i.e., even if we assume an intrinsic HB width of
zero,  $\langle \Delta V \rangle_{\rm HB}$ in the outer regions would
be significantly in excess of the observed values.
%This  is the reason why we rely on SDSS data in our analysis.

In summary, Draco has a remarkably thin blue HB showing no signs of variation
with radius.  Tidal models, however, predict a strong increase of HB thickness
with considered field size. While in the central regions, the HB width in the
best-fit tidal model still only marginally exceeds the values observed in
Draco, this excess is highly significant at larger radii. The large area
coverage offered by SDSS is needed to derive this result, and taking all
together, we feel save to conclude that even best-fit tidal models which are
able to reproduce the morphological parameters of Draco cannot correctly
predict its HB properties.

\section{Summary -- The Failure of the Tidal Model for Draco}
\label{sec:summary}
The dSph galaxies in the vicinity of the Milky Way are either strongly
dark-matter dominated, or they are the remnants of tidally disrupted
satellite galaxies and are seen today by terrestrial observers at
certain highly favorable viewing angles. This is inferred from the
observed stellar velocity dispersions, which are large with respect to
the luminous mass of the galaxy. A possible third alternative is a
modification to Newton's law of gravity (Sanders \& McGaugh 2002).
 
It was our attempt in this paper to construct an appropriate tidal model for
the Draco dSph galaxy.  Draco is particularly useful for testing
tidal models, because it is one of the best-studied satellite galaxies of our
Milky Way, and its compact and seemingly undistorted morphological appearance
as well as its thin HB in the CMD set tight
constraints on tidal models. Our effort to find a model for Draco that is able
to reproduce {\em all} properties of the galaxy known to date failed.

Our line of reasoning is the following. First, we consider the apparent lack
of `extra-tidal' stars in the vicinity of Draco to constrain the orbital
trajectory of the progenitor satellite galaxy.  Indeed, tidal models without a
noticeable number of `extra-tidal' stars are possible if the viewing angle to
the galaxy is close to a tangent to the orbital trajectory (see Figure
\ref{fig:trajectory}), which requires that the satellites follow very
eccentric orbits. Second, the observed compactness of Draco can be reproduced
if the satellite galaxy itself initially is small and compact. However, this
galaxy cannot be too compact, otherwise it will not be disrupted in the tidal
field of the Milky Way over a Hubble time. Third, rapid expansion of the tidal
debris after complete disruption limits the period during which the tidal
models resemble Draco to about $1\,$Gyr. A similar timespan can be derived
from looking at the line-of-sight velocity dispersion. Together, these
morphological and kinematic constraints severely limit the available parameter
space for tidal models, but do not render them completely implausible.
Finally, large field-of-view photometric data from the SDSS can be used to
fully exclude tidal models for Draco on the basis of its HB thickness. The
observed small variations in the position of stars in the CMD of Draco are
inconsistent with the predicted wide distribution in the tidal model
%especially at large radii
 due to distance modulus variations along the line of
sight. These models furthermore exhibit a strong increase of HB thickness with
increasing field size, while Draco exhibits no such variation. 

Altogether, we conclude that Draco {\em cannot} be the remnant a tidally
disrupted satellite galaxy. Instead, it most likely is strongly 
dark-matter-dominated as predicted by standard cosmological scenarios and,
increasingly, by observational evidence such as structural studies
(e.g., Odenkirchen et al.\ 2001a) and the radial dependence of the velocity
dispersion (see Kleyna et al.\ 2001, 2002). 
% Alternatively,
% modifications to the law of gravitation may be required (Sanders \&
% McGaugh 2002).

\acknowledgements{We thank J.\ S.\ Gallagher, P.\ Kroupa, M.\ 
  Odenkirchen, R.\ Scholz, and H.\ Zhao for stimulating discussions.
  We also thank the referee for many useful comments and suggestions.
  R.S.K. acknowledges support by the Emmy Noether Program of the
  Deutsche Forschungsgemeinschaft (DFG, grant KL1358/1) and subsidies
  from a NASA astrophysics theory program supporting the joint Center
  for Star Formation Studies at NASA-Ames Research Center, UC
  Berkeley, and UC Santa Cruz.
  
  Funding for the creation and distribution of the SDSS Archive has
  been provided by the Alfred P. Sloan Foundation, the Participating
  Institutions, the National Aeronautics and Space Administration, the
  National Science Foundation, the U.S. Department of Energy, the
  Japanese Monbukagakusho, and the Max Planck Society. The SDSS Web
  site is http://www.sdss.org/.
  
  The SDSS is managed by the Astrophysical Research Consortium (ARC)
  for the Participating Institutions. The Participating Institutions
  are The University of Chicago, Fermilab, the Institute for Advanced
  Study, the Japan Participation Group, The Johns Hopkins University,
  Los Alamos National Laboratory, the Max-Planck-Institute for
  Astronomy (MPIA), the Max-Planck-Institute for Astrophysics (MPA),
  New Mexico State University, University of Pittsburgh, Princeton
  University, the United States Naval Observatory, and the University
  of Washington.
  
  This research has made extensive use of NASA's Astrophysics Data
  System.}

\end{document}